\documentclass[11pt]{article}
\usepackage{graphicx}
\usepackage{amssymb}
\usepackage{graphics}
\usepackage{booktabs}
\usepackage{multirow}
\usepackage{caption}

\begin{document}

\title{Multi-mirror imaging optics for low-loss transport of divergent
neutron beams and tailored wavelength spectra}
\author{Oliver Zimmer \\
Institut Laue Langevin, 38042 Grenoble, France}
\maketitle

\begin{abstract}
\noindent A neutron optical transport system is proposed which comprises
nested short elliptical mirrors located halfway between two common focal
points M and M'. It images cold neutrons from a diverging beam or a source
with finite size at M by single reflections onto a spot of similar size at
M'. Direct view onto the neutron source is blocked by a central absorber
with little impact on the transported solid angle. Geometric neutron losses
due to source size can be kept small using modern supermirrors and distances
M-M' of a few tens of metres. Very short flat mirrors can be used in
practical implementations. Transport with a minimum of reflections remedies
losses due to multiple reflections that are common in long elliptical
neutron guides. Moreover, well-defined reflection angles lead to new
possibilities for enhancing the spectral quality of primary beams, such as
clear-cut discrimination of short neutron wavelengths or beam monochromation
using bandpass supermirrors. Multi-mirror imaging systems may thus
complement or even replace ordinary neutron guides, in particular at the
European Spallation Source. \bigskip

Keywords: neutron optics, supermirror, bandpass mirror, cold neutrons,
neutron EDM
\end{abstract}

\section{Introduction}

Physics with slow neutrons at research reactors and spallation sources can
hardly be imagined without devices for neutron transport. Thanks to the
invention of neutron guides \cite{Maier-Leibnitz/1963} intense beams have
become available over $100$ metres away from a neutron moderator. As a
result, many experiments can be performed simultaneously at a single source
and well shielded against radiation backgrounds. The physical principle of a
neutron guide is total reflection by a surface of a medium with a positive
neutron optical potential. For neutrons with wavelength $\lambda $ this
requires the normal component $k_{\bot }$ of the wave vector with modulus $%
k=2\pi /\lambda $ to stay below a critical value $k_{\bot \mathrm{c}}$
characteristic for the mirror material. Nickel with natural isotopic
composition is well suited, and enriched $^{58}$Ni even more so, for which $%
k_{\bot \mathrm{c}}$ is particularly large. For thermal and cold neutrons, $%
k\gg k_{\bot \mathrm{c}}$ corresponds to a small critical angle of
reflection that is given by%
\begin{equation}
\theta _{\mathrm{c}}\approx \tan \theta _{\mathrm{c}}=k_{\bot \mathrm{c}%
}/k=\kappa _{\mathrm{c}}\lambda .  \label{critical angle}
\end{equation}%
This defines a "mirror constant" $\kappa _{\mathrm{c}}$ with value $0.0173$
rad/nm for natural Ni.

Advances in performance have been achieved with the so-called supermirror.
This device relies on coherent wave superpositions in a superlattice formed
by a sequence of bilayers of two materials with very different neutron
optical potentials (most often Ni and Ti) \cite{Mezei/1976}. Broad-band
supermirrors became popular in neutron physics in the 1980's soon after Otto
Sch\"{a}rpf started a first large-scale production for a neutron scattering
instrument with wide-angle polarisation analysis \cite{Schaerpf/1991}. An
ideal device would extend by a factor $m$ the range $0<\kappa =k_{\bot
}/2\pi <\kappa _{\mathrm{c}}$ for which reflection from a simple mirror made
of natural Ni is nearly lossless. Correspondingly, the neutron flux at the
end of a long supermirror guide would scale with $m^{2}$ as the solid angle
increases. In real supermirrors, however, neutron losses occur due to
roughness of the layer interfaces and due to absorption. For simulations of
neutron transport efficiencies, one usually describes their reflectivity by%
\begin{equation}
R_{\mathrm{broad-band}}\left( \kappa \right) =\left\{ 
\begin{array}{cc}
1 & 0<\kappa \leq \kappa _{\mathrm{c}} \\ 
1-\frac{1-R_{\mathrm{bb}}}{\left( m-1\right) \kappa _{\mathrm{c}}}\left(
\kappa -\kappa _{\mathrm{c}}\right)  & \kappa _{\mathrm{c}}<\kappa \leq
m\kappa _{\mathrm{c}} \\ 
0 & \kappa >m\kappa _{\mathrm{c}}%
\end{array}%
\right. ,  \label{SM reflectivity}
\end{equation}%
which includes a linear drop from unity at $\kappa =\kappa _{\mathrm{c}}$
down to a value $R_{\mathrm{bb}}$ at $\kappa =m\kappa _{\mathrm{c}}$.
Commercially available supermirrors have reflectivity $R_{\mathrm{bb}%
}\approx 80$ \% at $m=4$. Hence, even if a neutron requires only a few
reflections with $\kappa $ in the range $\kappa _{\mathrm{c}}<\kappa \leq
m\kappa _{\mathrm{c}}$, its probability to arrive at the end of a long guide
may be quite low.

The concept of a ballistic neutron guide mitigates this problem \cite%
{Mezei/1997}. Here, the guide cross section first increases linearly with
the distance from the source, then stays constant in a second and usually
much longer section, and finally converges linearly back to its initial
size. The corresponding phase space transformations of the beam reduce the
total number of reflections. Moreover, the reduced beam divergence in the
straight guide section necessitates only a coating with lower $m$ which
costs less and has a higher reflectivity of typically $R_{\mathrm{bb}%
}\gtrsim 92$ \% at a standard $m=2$. A ballistic guide was first implemented
in the fundamental-physics beamline PF1B at the ILL \cite%
{Haese/2002,Abele/2006}. In the meantime, extensions of the concept have
become popular, including parabolically and elliptically shaped guides \cite%
{Schanzer/2004,Kleno/2012}.

It should be noted that neutron guides are not imaging devices and typically
transport neutrons by many reflections. This contrasts with grazing-angle
imaging systems that were first developed for x-rays, such as Wolter optics
which focus a parallel beam by double reflections off a parabolic mirror
combined with a hyperbolic or elliptical one. Such systems are also feasible
for slow neutrons \cite{Mildner/2011}, for which a microscope based on three
nested coaxial Wolter mirrors has been demonstrated \cite{Liu/2013}. Here a
mirror system is proposed for imaging neutrons under large solid angle from
an extended neutron source or the end of a neutron guide onto a target area
of similar size. Based on single reflections, the imperfect mirror
reflectivity in the range $\kappa _{\mathrm{c}}<\kappa \leq m\kappa _{%
\mathrm{c}}$ has a rather limited impact so that one may expect significant
gains with respect to long guides. Moreover, a well defined angle of
incidence on every mirror surface element offers interesting new
opportunities for tailoring and monochromation of primary beams.

\section{Neutron optical transport in an elliptical guide}

Let us first recall mathematical properties of an elliptical mirror and
point out why a long elliptical neutron guide is a non-imaging device. In a
plane with cartesian coordinates $x$ and $y$,%
\begin{equation}
\frac{x^{2}}{a^{2}}+\frac{y^{2}}{b^{2}}=1  \label{ellipse}
\end{equation}%
describes an ellipse with long and short axes $2a$ and $2b$, respectively.
The latter are related with the distance $2L$ between the two focal points
according to 
\begin{equation}
a^{2}=L^{2}+b^{2}.  \label{a}
\end{equation}%
It is well known that a single specular reflection off an ellipse maps a
straight ray emitted in any direction from the first focal point M located
at $\left( -L,0\right) $ onto the second focal point M' at $\left(
L,0\right) $. For cold neutrons with not excessively large wavelength, the
maximum reflection angle $\Theta =m\theta _{\mathrm{c}}$ does not exceed a
few degrees even for a large-$m$ supermirror. Elliptical neutron mirrors are
therefore truncated to $\left\vert x\right\vert \leq l<L$, and guides
usually "narrow" ($a\gg b$) and "long" ($l\gg L-l$).

It is easy to see that for a long elliptical guide the mapping between focal
points does not lead to an optical image of an extended source. As
illustrated in Fig.\ $1$, a ray starting at $x=-L$ with a lateral offset $%
\Delta y$ and reflected at a point $\left( X,Y\right) $ arrives at $x=L$
with an offset of%
\begin{equation}
\Delta y^{\prime }\approx \frac{L-X}{L+X}\Delta y.  \label{magnification}
\end{equation}%
A long mirror reflects within a wide range of $X$ and hence superposes
images with widely varying magnifications $\Delta y^{\prime }/\Delta y$.
Reflections at $X>0$ are focusing, $\Delta y^{\prime }<\Delta y$, as shown
for example with the reflection point $\left( X_{3},Y_{3}\right) $. They
lead to an increase of flux density at M'. In the opposite case, $X<0$
[e.g., for reflection at $\left( X_{1},Y_{1}\right) $], the mirror magnifies
the source, thus reducing the corresponding flux density. The latter case
represents a vast majority of neutrons in a long guide, because the ratio of
ranges of unidimensional angular acceptance, $\Delta \xi \left( X<0\right)
/\Delta \xi \left( X>0\right) \approx \sqrt{\left( L+l\right) /\left(
L-l\right) }-1$, is usually much larger than unity and enters squared in the
solid angle. The spatial intensity distribution at M' thus becomes widened
and the flux density reduced with respect to M. In addition, for a
significant fraction of neutrons the magnification becomes so large that
they impinge at least a second time on the mirror [such as the ray from $%
\left( X_{4},Y_{4}\right) $ to $\left( X_{5},Y_{5}\right) $]. This leads not
only to increased losses due to imperfect mirror reflectivity, but also to
complex structures in the divergence distribution as observed in Monte-Carlo
simulations \cite{Cussen/2013}. Indeed, these authors concluded that
"transport of neutrons by realistic elliptic guides usually involves many
reflections, contrary to the usual expectations". They also state that
multiple reflections become increasingly common with increasing guide
length. A mirror system based on single neutron reflections may therefore
lead to improved beam transport efficiency. The next section discusses such
a system.

\begin{figure}[tbp]
\centering
\includegraphics[width=0.82\textwidth]{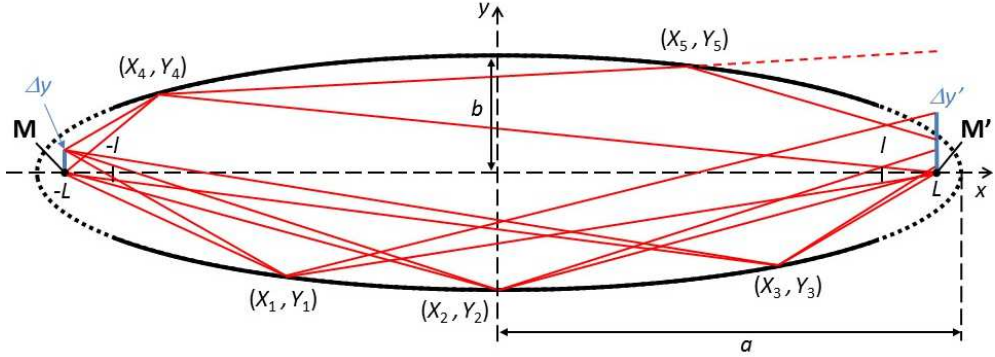}
\caption{"Imaging" in a long elliptical reflector. The cases discussed in
the text are shown, as well as the symmetric case $X_{2}=0$, for which the
reflection angles are minimised and $\Delta y^{\prime }/\Delta y=1$.}
\label{fig:figure1}
\end{figure}

\section{Elliptical multi-mirror system}

The proposed neutron optical system is shown in Fig.$\ 2$. It consists of
nested short elliptical mirrors with common length $2l$. They are located in
the plane halfway between M and M' which are the common focal points of all
ellipses. A central absorber blocks the direct view from M' to M. According
to Eq.\ \ref{magnification} and in contrast to a single long mirror, one
obtains a well-defined image at M' which is weakly blurred due to
magnifications being limited to a narrow range $1\pm 2l/L$. Focusing is also
possible, by placing elliptical mirrors closer to the point M'. However,
this reduces the solid angle subtended from the point M. Compared to a guide
starting close to a source, mirrors situated at a large distance (i.e., $L$)
are much less exposed to damaging radiation. A clear spatial separation from
the source environment also facilitates independent maintenance of
neutron-optical and in-pile components, including exchange of beam tubes.

The mirror system may be implemented either with toroidal or with
translational, "planar" symmetry. In the former case, the device is
rotationally symmetric about the $x$ axis; each mirror surface is a section
of an ellipsoid of revolution ($x$ and $y$ in Eq.\ \ref{ellipse} are then
taken as the axial and radial components in cylindrical coordinates,
respectively). A single reflection transports a neutron from M to M'. In the
planar case, the mirrors have a local translational symmetry along the $z$
axis (taken together with $x$ and $y$ as Cartesian coordinates). Neutrons
undergo a single reflection with momentum transfer in the $x,y$ plane, but
additional mirrors are needed to refocus or guide them in the orthogonal $z$
direction. For instance, adding a second planar multi-mirror optics rotated
by $90$ degrees about the $x$ axis creates a "double planar" system in which
two imaging reflections occur, one for each transverse dimension.

In contrast to grazing-angle optics for x-rays, toroidal mirrors for imaging
slow neutrons over long distances may exhibit significant chromatic
aberrations due to gravity. These also occur in the aforementioned double
planar system, obviously without affecting reflections on vertical mirrors.
Gravity has practically no relevant influence in a hybrid system consisting
of planar multi-mirror optics combined with simple or ballistic reflective
boundaries in the vertical dimension $z$. These mirrors, projected onto the $%
x,y$ plane, must cover the whole area defined by the outermost rays in Fig.\ 
$2$. Hybrid systems seem best suited when a large beam divergence can be
accepted only in one dimension. For instance, neutron scattering instruments
possessing a detector array with high unidimensional angular resolution
require a tight collimation within the scattering plane. An anisotropic beam
divergence may also originate from the source due to moderator shape and/or
geometrical constraints on beam tubes. Another interesting application of
hybrid systems is large-angle deviation of beams with low-to-medium
divergence, given that space restrictions for instruments at reactor and
spallation neutron sources are often an issue. In this case one uses only
half of the device shown in Fig.\ $2$.

\begin{figure}[tbp]
\centering
\includegraphics[width=0.9\textwidth]{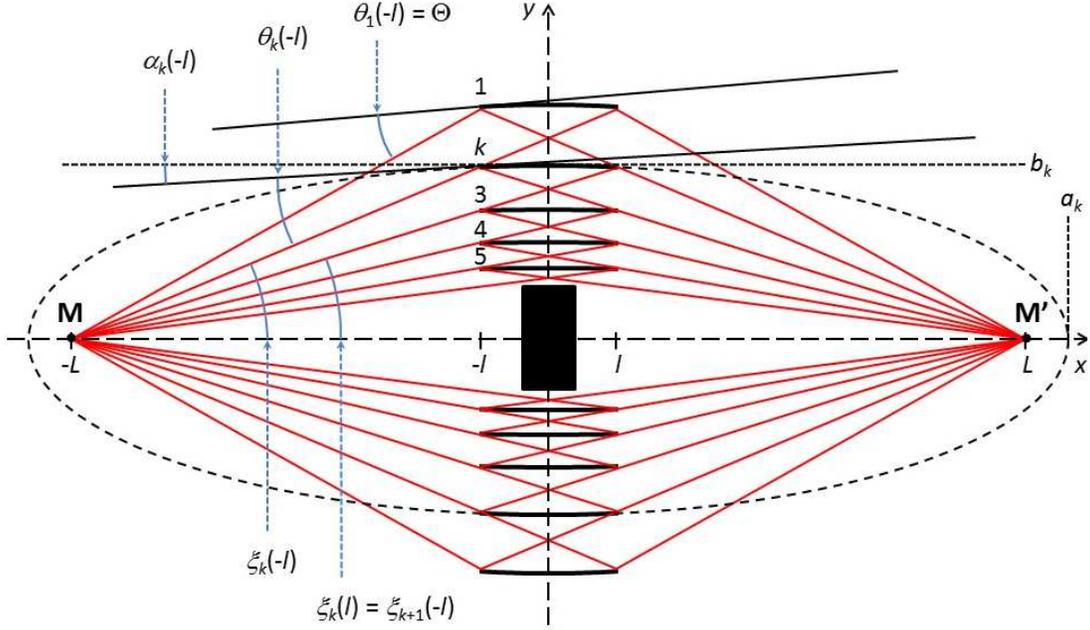}
\caption{Schematic of the nested elliptical multi-mirror system. For the
construction see text.}
\label{fig:figure2}
\end{figure}

For the mathematical construction let us define $\xi _{k}\left( x\right) $
as the angle between the axis connecting the focal points and a straight
line from M to a point $\left( x,y_{k}\right) $ on the ellipse with index $k$%
, as shown in Fig.\ $2$. The angle $\theta _{k}\left( x\right) $ of
reflection at $\left( x,y_{k}\right) $ then follows from%
\begin{equation}
\xi _{k}\left( x\right) =\theta _{k}\left( x\right) +\alpha _{k}\left(
x\right) =\arctan \left( \frac{b_{k}\sqrt{L^{2}+b_{k}^{2}-x^{2}}}{\sqrt{%
L^{2}+b_{k}^{2}}\left( L+x\right) }\right) ,  \label{xi}
\end{equation}%
where the angle $\alpha _{k}\left( x\right) $ is related to the slope of the
tangent to the ellipse according to%
\begin{equation}
\frac{d\left\vert y_{k}\left( x\right) \right\vert }{dx}=\frac{-b_{k}x}{%
\sqrt{L^{2}+b_{k}^{2}}\sqrt{L^{2}+b_{k}^{2}-x^{2}}}=\tan \alpha _{k}\left(
x\right) .  \label{alpha(x)}
\end{equation}%
Notice that all angles are taken as positive except $\alpha _{k}\left(
x\right) $, which becomes negative for $x>0$. These definitions provide a
consistent common description for the mirror systems with toroidal and
planar symmetry. The angle subtended at a focal point by an elliptical
section between $x=-l$ and $x=l$ is given by%
\begin{equation}
\Delta \xi _{k}=\xi _{k}\left( -l\right) -\xi _{k}\left( l\right) =2\alpha
_{k}\left( -l\right) .  \label{delta-xi}
\end{equation}%
For a point source at M this defines a range of angles that the mirror $k$
can accept. Using Eqs.\ \ref{xi} and \ref{alpha(x)}, the angles of
reflection for this mirror follow from%
\begin{equation}
\tan \theta _{k}\left( x\right) =\frac{b_{k}}{L}\frac{\sqrt{L^{2}+b_{k}^{2}}%
}{\sqrt{L^{2}+b_{k}^{2}-x^{2}}}.  \label{theta(x)}
\end{equation}%
Hence, a smaller value of $l$ produces a smaller spread of reflection angles.

The nested ellipses can be constructed as follows. Note first that 
\begin{equation}
y_{k}\left( l\right) =b_{k}\frac{\sqrt{L^{2}+b_{k}^{2}-l^{2}}}{\sqrt{%
L^{2}+b_{k}^{2}}},  \label{y1}
\end{equation}%
according to Eq.\ \ref{ellipse} applied to the ellipse $k$. The straight
line connecting the focal point M with the rear end point $\left(
l,y_{k}\right) $ of the mirror $k$ is required to coincide with the front
end point $\left( -l,y_{k+1}\right) $ of the mirror $k+1$. This provides the
condition [note that $y_{k+1}\left( -l\right) =y_{k+1}\left( l\right) $] 
\begin{equation}
y_{k+1}\left( l\right) =y_{k}\left( l\right) -2l\tan \xi _{k}\left( l\right)
=b_{k}\frac{\sqrt{L^{2}+b_{k}^{2}-l^{2}}\left( L-l\right) }{\sqrt{%
L^{2}+b_{k}^{2}}\left( L+l\right) },  \label{y-recursive}
\end{equation}%
where the right side follows from Eqs.\ \ref{xi} and \ref{y1}. The mirrors
are defined recursively by incrementing $k$ in unit steps in Eqs.\ \ref{y1}
and \ref{y-recursive}. The outermost ellipse is defined by a largest
reflection angle $\Theta $ that occurs at its ends, $\theta _{1}\left( \pm
l\right) =\Theta $. This leads us to%
\begin{equation}
b_{1}^{2}=-\frac{1}{2}L^{2}\left( 1-\tan ^{2}\Theta \right) +L\sqrt{\frac{1}{%
4}L^{2}\left( 1-\tan ^{2}\Theta \right) ^{2}+\left( L^{2}-l^{2}\right) \tan
^{2}\Theta }.  \label{b}
\end{equation}%
Table $1$ lists parameters of the mirrors of a representative assembly for
neutron transport to a position $30$ m away from a source.

The total number of mirrors $N$ corresponds to the number of recursions. For
the toroidal case, $N$ counts the rings of mirrors, while for the planar
system it counts the mirrors on one side of the central absorber. In the
limit $l/L\rightarrow 0$ one can estimate $N$ as follows without actually
performing the recursion. Using Eqs.\ \ref{y1} and \ref{y-recursive}, the
channel width between the mirrors $k$ and $k+1$ is given by%
\begin{equation}
W_{k,k+1}=y_{k}\left( l\right) -y_{k+1}\left( l\right) \approx \frac{2lb_{k}%
}{L},
\end{equation}%
where the expression on the right side is valid with a relative error of
order $l/L$. For $l\ll L$ we may also write $W_{k,k+1}\approx b_{k}-b_{k+1}$%
. Hence,%
\begin{equation}
b_{k+1}\approx b_{k}-\frac{2lb_{k}}{L},
\end{equation}%
so that%
\begin{equation}
b_{N}=\left( 1-\frac{2l}{L}\right) ^{N-1}b_{1}.
\end{equation}%
The number of mirrors needed to cover the range of angles defined by the
parameters $b_{1}$ and $b_{N}$ of the outer- and innermost ellipses,
respectively, is thus given by%
\begin{equation}
N=\frac{\ln \frac{b_{N}}{b_{1}}}{\ln \left( 1-\frac{2l}{L}\right) }+1\approx 
\frac{L}{2l}\ln \frac{b_{1}}{b_{N}}+1.  \label{N}
\end{equation}

\begin{table}[tbp] \centering%
$%
\begin{tabular}{l|lllllll}
$k$ & $y_{k}\left( \pm l\right) $ $\left[ \mathrm{m}\right] $ & $b_{k}$ $%
\left[ \mathrm{m}\right] $ & $a_{k}$ $\left[ \mathrm{m}\right] $ & $\xi
_{k}\left( -l\right) $ $\left[ ^{0}\right] $ & $\theta _{k}\left( \pm
l\right) $ $\left[ ^{0}\right] $ & $\theta _{k}\left( 0\right) $ $\left[ ^{0}%
\right] $ & $m_{k}$ \\ \hline
$-2$ & $2.\,\allowbreak 307\,3$ & $2.\,\allowbreak 312\,3$ & $%
15.\,\allowbreak 177\,2$ & $9.\,\allowbreak 359$ & $8.\,\allowbreak 782$ & $%
8.\,\allowbreak 763$ & $10$ \\ 
$-1$ & $2.\,\allowbreak 018\,9$ & $2.\,\allowbreak 023\,3$ & $%
15.\,\allowbreak 135\,8$ & $8.\,\allowbreak 206$ & $\allowbreak
7.\,\allowbreak 699$ & $7.\,\allowbreak 682$ & $\allowbreak 8.\,\allowbreak
80$ \\ 
$0$ & $1.\,\allowbreak 766\,5$ & $1.\,\allowbreak 770\,4$ & $%
15.\,\allowbreak 104\,1$ & $\allowbreak 7.\,\allowbreak 192$ & $\allowbreak
6.\,\allowbreak 746$ & $6.\,\allowbreak 731$ & $7.\,\allowbreak 70$ \\ 
$1$ & $1.\,\allowbreak 545\,7$ & $\allowbreak 1.\,\allowbreak 549\,1$ & $%
15.\,\allowbreak 079\,8$ & $6.\,\allowbreak 300$ & $5.\,\allowbreak 909$ & $%
5.\,\allowbreak 896$ & $6.\,\allowbreak 74$ \\ 
$2$ & $1.\,\allowbreak 352\,5$ & $\allowbreak 1.\,\allowbreak
355\,5\allowbreak $ & $15.\,\allowbreak 061\,1$ & $5.\,\allowbreak 518$ & $%
5.\,\allowbreak 175$ & $5.\,\allowbreak 163$ & $\allowbreak 5.\,\allowbreak
90$ \\ 
$3$ & $1.\,\allowbreak 183\,4$ & $1.\,\allowbreak 186\,0$ & $%
15.\,\allowbreak 046\,8$ & $\allowbreak 4.\,\allowbreak 832$ & $\allowbreak
4.\,\allowbreak 531$ & $4.\,\allowbreak 521$ & $5.\,\allowbreak 16$ \\ 
$4$ & $1.\,\allowbreak 035\,5$ & $1.\,\allowbreak 037\,8$ & $%
15.\,\allowbreak 035\,9$ & $\allowbreak 4.\,\allowbreak 230$ & $\allowbreak
3.\,\allowbreak 967$ & $3.\,\allowbreak 958$ & $\allowbreak 4.\,\allowbreak
51$ \\ 
$5$ & $\allowbreak 0.906\,1\,$ & $0.908\,1$ & $15.\,\allowbreak 027\,5$ & $%
\allowbreak 3.\,\allowbreak 703$ & $\allowbreak 3.\,\allowbreak 472$ & $%
\allowbreak 3.\,\allowbreak 464$ & $\allowbreak 3.\,\allowbreak 95$ \\ 
$6$ & $0.792\,8$ & $0.794\,6$ & $15.\,\allowbreak 021$ & $\allowbreak
3.\,\allowbreak 241$ & $\allowbreak 3.\,\allowbreak 039$ & $\allowbreak
3.\,\allowbreak 032$ & $\allowbreak 3.\,\allowbreak 46$ \\ 
$7$ & $0.693\,7$ & $0.695\,2$ & $15.\,\allowbreak 016\,1$ & $\allowbreak
2.\,\allowbreak 837$ & $\allowbreak 2.\,\allowbreak 660$ & $\allowbreak
2.\,\allowbreak 654$ & $\allowbreak 3.\,\allowbreak 02$ \\ 
$8$ & $0.607\,0$ & $0.608\,3$ & $15.\,\allowbreak 012\,3$ & $\allowbreak
2.\,\allowbreak 483$ & $2.\,\allowbreak 328$ & $2.\,\allowbreak 322$ & $%
\allowbreak 2.\,\allowbreak 65$ \\ 
$9$ & $\allowbreak 0.531\,1$ & $0.532\,3$ & $15.\,\allowbreak 009\,4$ & $%
2.\,\allowbreak 173$ & $2.\,\allowbreak 037$ & $2.\,\allowbreak 032$ & $%
2.\,\allowbreak 32$ \\ 
$10$ & $0.464\,7$ & $\allowbreak 0.465\,8$ & $15.\,\allowbreak 007\,2$ & $%
\allowbreak 1.\,\allowbreak 901$ & $1.\,\allowbreak 782$ & $1.\,\allowbreak
779$ & $2.\,\allowbreak 03$ \\ 
$11$ & $0.406\,6$ & $0.407\,5$ & $15.\,\allowbreak 005\,5$ & $%
1.\,\allowbreak 664$ & $1.\,\allowbreak 560$ & $\allowbreak 1.\,\allowbreak
556$ & $1.\,\allowbreak 78$ \\ 
$12$ & $0.355\,8$ & $0.356\,6$ & $15.\,\allowbreak 004\,2$ & $%
1.\,\allowbreak 456$ & $1.\,\allowbreak 365$ & $1.\,\allowbreak 362$ & $%
1.\,\allowbreak 55$ \\ 
$13$ & $0.311\,3$ & $0.312\,0$ & $15.\,\allowbreak 003\,2$ & $%
1.\,\allowbreak 274$ & $1.\,\allowbreak 194$ & $1.\,\allowbreak 192$ & $%
1.\,\allowbreak 36$ \\ 
$14$ & $0.272\,4$ & $0.273\,0$ & $15.\,\allowbreak 002\,5$ & $%
1.\,\allowbreak 115$ & $1.\,\allowbreak 045$ & $1.\,\allowbreak 043$ & $%
1.\,\allowbreak 19$ \\ 
$15$ & $0.238\,4\allowbreak $ & $0.238\,9$ & $15.\,\allowbreak 001\,9$ & $%
\allowbreak 0.975$ & $0.914$ & $\allowbreak 0.912$ & $1.\,\allowbreak 04$ \\ 
$16$ & $0.208\,6$ & $\allowbreak 0.209\,0$ & $15.\,\allowbreak 001\,5$ & $%
\allowbreak 0.854$ & $\allowbreak 0.800$ & $0.798$ & $\allowbreak 0.91$ \\ 
$17$ & $0.182\,5$ & $0.182\,9$ & $15.\,\allowbreak 001\,1$ & $0.747$ & $%
0.700 $ & $0.699$ & $0.80$ \\ 
$18$ & $0.159\,7$ & $0.160\,0$ & $15.\,\allowbreak 000\,9$ & $\allowbreak
0.653$ & $0.613$ & $0.611$ & $0.70$%
\end{tabular}%
$%
\caption{Mirror assembly with $L=15\ \mathrm{m}$ and $l=1\ \mathrm{m}$. 
The values $m_{k}$ were calculated for a common $\tilde{\lambda}=0.886\ \mathrm{nm}$ using 
Eq.\ \ref{cutoff condition}. Mirrors up to $m=7$ are commercially
available, i.e., $k\geq 1$ in the list. Notice the decrease of channel widths  from almost $30\ 
\mathrm{cm}$ down to $2\ \mathrm{cm}$. 
A criterion for the maximum useful $k$ is the 
solid angle that the central absorber sketched in Fig.\ $2$ must cover for a 
tolerable radiation background near M'. For example, in a beam deviator (see text),
mirrors in the ranges $k=3-7$ and $3-15$ would transport divergences $m=1$ and $2$, 
and deviate the beam by $7.2$ and $5.6$ degrees, respectively, from its original direction.}%
\label{TableKey copy(1)}%
\end{table}%

As a new feature for neutron optical transport systems (and in contrast to
neutron guides), the multi-mirror optics discriminates short-wavelength
neutrons. These are often a nuissance and lead for instance to second-order
Bragg diffraction in crystal monochromators. While a Bragg filter made of
compressed crystallites of a weakly absorbing material can remove this
spectral component, some useful neutrons are lost as well. Short-wavelength
neutrons also generate scattering backgrounds that, due to the $1/v$ law for
neutron absorption, require thicker shielding. A multi-mirror imaging system
supersedes Bragg filters and improves background conditions. Due to
reflection angles in a narrow range about a finite value, each mirror
provides a sharp cutoff. Moreover, the spectrum reflected from every mirror
can be tailored to the same cutoff wavelength $\tilde{\lambda}$, which is
defined by Eq.\ \ref{critical angle} (replacing $\kappa _{\mathrm{c}}$ by $%
m_{k}\kappa _{\mathrm{c}}$ to account for the extended reflectivity of
supermirrors) and the geometric definition of the reflection angle on the
mirror $k$. It thus satisfies the relationships 
\begin{equation}
m_{k}\kappa _{\mathrm{c}}\tilde{\lambda}=\frac{b_{k}}{L}.
\label{cutoff condition}
\end{equation}%
In practice, one could first select a value for $\tilde{\lambda}$, then
choose an $m$-value for the outermost mirror and finally determine the
values $m_{k}$ of all the other mirrors, while keeping constant the ratios $%
m_{k}/b_{k}$ (this has been done in Table $1$). Obviously, a smaller spread
of reflection angles on each mirror (see Eq.\ \ref{theta(x)}) leads to a
better definition of $\tilde{\lambda}$. This requires the mirrors to be
shorter (and thus a larger number of them according to Eq.\ \ref{N}).

Next we discuss neutron losses due to finite size of the source at M. We
call them "geometrical" for distinction from losses due to imperfect mirror
reflectivity. For simplicity let us consider the case of the planar system.
As sketched in the upper part of Fig.\ $3$, neutrons starting from the point 
$\left( -L,\Delta y>0\right) $ with a positive velocity component $\dot{y}$
may fly through the mirror channel $\left[ k,k+1\right] $ without any
reflection. For a starting point $\left( -L,\Delta y<0\right) $, shown in
the lower part of Fig.\ $3$, neutrons may become reflected twice. For each
channel $\left[ k,k+1\right] $ one can quantify a transport loss associated
with these effects. For a positive lateral displacement $\Delta y>0$ (with $%
\dot{y}>0$) this is given by 
\begin{equation}
\Gamma _{k}\left( \Delta y>0\right) =\frac{\zeta _{\mathrm{c}}-\zeta
_{k+1}\left( -l\right) }{\zeta _{k}\left( -l\right) -\zeta _{k+1}\left(
-l\right) },
\end{equation}%
where $\zeta _{\mathrm{c}}=\zeta _{k}\left( l\right) $ is a critical angle
that separates a range of angles $\zeta $ for which no reflections occur, $%
\zeta _{\mathrm{c}}>\zeta >\zeta _{k+1}\left( -l\right) $, from the range $%
\zeta _{k}\left( -l\right) >\zeta >\zeta _{\mathrm{c}}$ corresponding to the
desired single reflections. For $\Delta y<0$ the fractional transport loss
is given by%
\begin{equation}
\Gamma _{k}\left( \Delta y<0\right) =\frac{\zeta _{k}\left( -l\right) -\zeta
_{\mathrm{c}}}{\zeta _{k}\left( -l\right) -\zeta _{k+1}\left( -l\right) },
\end{equation}%
where $\zeta _{k}\left( -l\right) >\zeta >\zeta _{\mathrm{c}}$ is the range
of angles for which double reflections are possible. Most of the angles can
be directly read from Fig.\ $3$, whereas $\zeta _{\mathrm{c}}$ in the case $%
\Delta y<0$ needs to be calculated. For $l\ll L$ one obtains, independent of
the sign of displacement, 
\begin{equation}
\Gamma _{k}\left( \Delta y\right) \approx \frac{\left\vert \Delta
y\right\vert }{b_{k}}.  \label{gamma}
\end{equation}%
The innermost mirror channel thus has the largest losses. For a source of a
given size, geometrical losses decrease when the mirror system is enlarged.
Keeping the ratios $b_{k}/L$ and $l/L$ constant, the angular acceptance
stays approximately constant (exactly so for a point source), while $\Gamma
_{k}$ from Eq.\ \ref{gamma} decreases with increasing $b_{k}$. For a system
with given parameters $L$, $l$, $N$, and $m_{k}\kappa _{\mathrm{c}}$
designed for a common $\tilde{\lambda}$, we have $b_{k}\propto \tilde{\lambda%
}$ according to Eq.\ \ref{cutoff condition}. Hence, a system for cold
neutrons is shorter than an equivalent system for thermal neutrons with same
geometrical losses. For too large $L$, however, the performance may become
limited by mirror waviness. Note finally that $\Gamma _{k}$ does not depend
on $l$ (in the limit $l\ll L$). Systems consisting of very short, flat
mirrors locally approximating the elliptical surfaces are thus a viable
possibility.

\begin{figure}[tbp]
\centering
\includegraphics[width=0.82\textwidth]{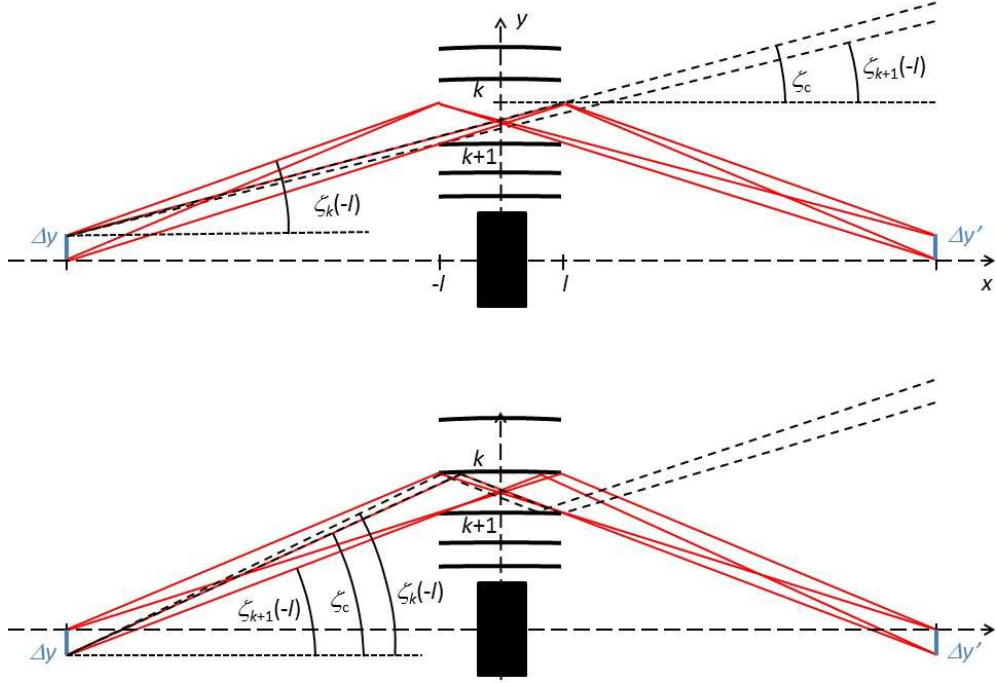}
\caption{Neutron transport losses due to no reflection (upper figure) or
double reflection (lower figure) in the mirror channel $\left[ k,k+1\right] $%
. The angular range of neutrons that miss the target region is defined by
the dashed rays.}
\label{fig:figure3}
\end{figure}

A first estimate of the efficiency of a mirror system involves the solid
angle for neutron transport. For a sufficently large toroidal system it is
characterised by the angles $\xi _{1}\left( -l\right) $ and $\xi _{N}\left(
l\right) $ of the outer- and innermost rays shown in Fig.\ $2$, i.e.%
\begin{equation}
\Delta \Omega =2\pi \left[ \cos \xi _{N}\left( l\right) -\cos \xi _{1}\left(
-l\right) \right] .  \label{acceptance solid}
\end{equation}%
For a planar system with $2N$ nested mirrors one may define a unidimensional
angular acceptance%
\begin{equation}
\Delta \xi =\xi _{1}\left( -l\right) -\xi _{N}\left( l\right)
=2\sum_{i=1}^{N}\alpha _{i}\left( -l\right) .  \label{acceptance linear}
\end{equation}%
Determination of the transport solid angle also requires knowledge of the
angular acceptance in the orthogonal transverse direction. For calculation
of the transport efficiency, Eqs.\ \ref{acceptance solid} and \ref%
{acceptance linear} still need to be corrected for imperfect mirror
reflectivities and for the geometrical losses described before. The simple
kinematics of single reflections facilitates numeric calculations without
simulating trajectories. For a known source this should provide fairly
accurate predictions of neutron fluxes.

\section{Wavelength selective neutron transport optics}

When neutrons within a limited range of wavelengths are needed for a
specific application, the multi-mirror systems described before can be
equipped with bandpass supermirrors for transport of the desired spectrum.
In particular, this includes quasi-monochromatic spectra peaked at some
value with a spread of a few percent or less. For a discussion of the
principle let us describe their reflectivity approximately by%
\begin{equation}
R_{\mathrm{bandpass}}\left( \kappa \right) =\left\{ 
\begin{array}{cc}
1 & 0<\kappa \leq \kappa _{\mathrm{c}} \\ 
0 & \kappa _{\mathrm{c}}<\kappa \leq \left( m-\delta \right) \kappa _{%
\mathrm{c}} \\ 
R_{\mathrm{bp}} & \left( m-\delta \right) \kappa _{\mathrm{c}}<\kappa \leq
m\kappa _{\mathrm{c}} \\ 
0 & \kappa >m\kappa _{\mathrm{c}}%
\end{array}%
\right. ,  \label{BPF reflectivity}
\end{equation}%
where the bandpass reflectivity $R_{\mathrm{bp}}$ is separated from a region
of total reflectivity at low $\kappa $. A bandpass supermirror requires
fewer layers than a broad-band supermirror with same $m$ value. It therefore
has less absorption and can be produced with less roughness. This leads to
higher reflectivity, $R_{\mathrm{bp}}>R_{\mathrm{bb}}$. For instance, using $%
4664$ single layers, Hino and coworkers have already achieved a bandpass
mirror peaking at $m=6.1$ with $79$ \% reflectivity and a width of $6$ \%
(FWHM) \cite{Hino/2010}. On the other hand, the best broad-band supermirrors
require $3$ to $4$ times as many layers and provide reflectivities up to
only $R_{\mathrm{bb}}\approx 63$ \%. Bandpass mirrors can thus enhance the
efficiency and reduce the cost for neutron transport of restricted
wavelengths. Following the arguments from the previous section for a common
cutoff wavelength $\tilde{\lambda}$ and broad-band supermirrors, one can
tune the individual bandpass mirrors $k$ to a common characteristic
wavelength (e.g., the shortest) within the band to be transmitted, e.g., $%
\lambda _{0}=b_{k}/Lm_{k}\kappa _{\mathrm{c}}$, with $\delta _{k}$
sufficiently wide to cover the band. If a multi-mirror design includes
innermost mirrors that reflect neutrons with wavelength $\lambda _{0}$ at $%
m_{k}<1$ (see Table $1$), simple $m=1$ coatings can be used since bandpass
reflectivity is no longer available.

If spectral purity is a design priority, one may exclude the mirrors with $%
m_{k}<1$, which reflect neutrons with wavelengths below and above $\lambda
_{0}$. In any case these cover only a small solid angle. Mirrors with $%
m_{k}>1$ reflect neutrons totally at wavelengths $\lambda \geq b_{k}/L\kappa
_{\mathrm{c}}>\lambda _{0}$, with the highest cutoff for the outermost
mirror. The multi-mirror assembly should be designed to reflect the desired
wavelength band using the minimum possible widths $\delta _{k}$. This
requires the spread of reflection angles on each mirror to be reduced to a
minimum and therefore also a large $L$, as this spread depends on the ratios 
$l/L$ and $D/L$, where $D$ denotes the size of the source. For a narrow
spectrum the width $\Delta \lambda /\lambda _{0}$ sets a scale for both
ratios (Ref.\ \cite{Masalovich/2013} discusses limits to the achievable
minimum bandwidth of neutron supermirrors). When shorter mirrors are chosen
for minimising the ratio $l/L$, the filling factor of substrates in the
channels is increased. Using a design with longer mirrors (staying within
the limits set by the desired quality of the image at M') one can subdivide
them into sections with individually adapted coatings of smaller bandwidth.
Another option is to employ a stack of very short, flat mirrors deposited on
highly transmissive substrates. Interestingly, because $\kappa _{\mathrm{c}}$
was defined for an interface of Ni to vacuum, reflection by supermirrors
within a substrate matrix with a positive neutron optical potential reduces
the contrast to the Ni layers and therefore reduces the range for total
reflection, $\kappa <\kappa _{\mathrm{c}}$\footnote{%
Note that this property of densely stacked mirrors was already pointed out
in Ref.\ \cite{Mook/1988} as a means for complete suppression of total reflection of one
neutron spin state in a polarising device. Indeed, magnetised Fe(20\%Co)/Al
supermirror layers and Si substrates have well matched refractive indices
for this state, enabling very high polarisation. Ref.\ \cite{Petukhov/2016}
reports first results of an adaptation of the concept to Fe/SiN$_{x}$ supermirrors.}. Another way to shift the total
reflection background to longer wavelengths is to replace the Ni layers of
the innermost mirrors by a material with a lower neutron optical potential.

A nested-mirror system equipped with bandpass mirrors may increase beam
intensities and reduce backgrounds. For some applications it supersedes
separate beam-tailoring devices such as neutron velocity selectors or
monochromators. One can imagine a plethora of new options for neutron
instrumentation, such as an exchangeable set of several planar systems for
different wavelengths, or angular-encoded multiplexed beams. As mentioned
above, gravity leads to chromatic aberrations in toroidal and double planar
multi-mirror systems, but has no practical influence in hybrid optics with
simple or ballistic mirrors for vertical reflections. For a double-planar
system for monochromatic neutrons, gravity can be easily compensated by a
proper choice of ellipse parameters (which then become different for the
mirrors above and below the central absorber).

The initial motivation for the present work\ was improving a method for
production of ultracold neutrons (UCNs) that employs a cold neutron beam in
superfluid helium held at temperatures below $1$ K \cite{Golub/1977}. This
method relies on conversion of neutrons with wavelengths in a narrow range
about $0.89$ nm, which can lose most of their energy by single-phonon
emission in the helium\footnote{%
Conversion of neutrons to UCNs with energies up to about $230\ \mathrm{neV}$
(which can be trapped in a superfluid-helium filled Be bottle) requires cold
neutrons with energies between $1.0289\ \mathrm{meV}$ and $1.0536\ \mathrm{%
meV}$ \cite{Yoshiki/2003}, i.e., neutron wavelengths between $0.8812\ 
\mathrm{nm}$ and $0.8917\ \mathrm{nm}$. The multi-mirror system thus needs
to reflect a minimum bandwidth of $\Delta \lambda /\lambda =1.19\%$ around $%
\lambda _{0}=0.8865\ \mathrm{nm}$.}. UCNs can be trapped in "neutron
bottles" made of neutron optical and/or magnetic and gravitational
potentials and are used for various applications in fundamental physics \cite%
{Dubbers/2011,Musolf/2008,Abele/2008} (for techniques and experiments using
UCNs see the books \cite{Golub/1991,Ignatovich/1990}). Owing to the
vanishing absorption cross section of $^{4}$He and a small probability for
scattering UCNs back to higher energies, a converter with reflective
boundaries can accumulate a high density of UCNs. If this is to be achieved
in storage chambers of size comparable with the converter, a
superfluid-helium UCN source at the end of a cold-neutron guide \cite%
{Golub/1977,Ageron/1978,Golub/1983,Kilvington/1987,Huffman/2000,Baker/2003,Zimmer/2007,Zimmer/2010,Zimmer/2011,Piegsa/2014,Zimmer/2015,Schmidt-Wellenburg/2015,Leung/2016}%
, can be competitive with "in-pile" implementations close to a cold-neutron
source \cite{Yoshiki/1994,Masuda/2002,Serebrov/2010,Masuda/2012}. The latter
have however higher UCN production rates and can hence provide higher
steady-state UCN fluxes and total UCN numbers in large vessels. Advantages
of an "in beam" source position far from strong radiation fields include
absence of nuclear licencing, reduced cooling power, easy access, and
short-distance UCN transport. A helium converter may even be implemented as
an integral part of the experiment as in a planned search for a neutron
electric dipole moment at the SNS \cite{EDM@SNS/2004}. The former CryoEDM
collaboration has already demonstrated polarised cold neutron conversion to
polarised UCNs \cite{Grinten/2009}. A multi-mirror system for monochromatic
and polarised neutrons would thus be particularly efficient for this
application. Monte-Carlo simulations for a system optimised for the small
liquid hydrogen moderators foreseen at the European Spallation Source \cite%
{TDR-ESS/2012} will be presented elsewhere \cite{Kepka/2016}.

\end{document}